\documentclass[%
preprint,
superscriptaddress,
showpacs,
preprintnumbers,
 bibnotes,
 amsmath,amssymb,
 aps,
prb,
longbibliography
]{revtex4-1}

\usepackage{graphicx}
\usepackage{dcolumn}
\usepackage{bm}
\usepackage{color}

\begin{document}


\title{Large Magnetoresistance in Compensated Semimetals TaAs$_2$ and NbAs$_2$}
\author{Zhujun Yuan,$^{1}$ Hong Lu,$^{1}$ Yongjie Liu,$^{3}$ Junfeng Wang,$^{3}$ Shuang Jia$^{1,2\S}$}
\affiliation{$^1$International Center for Quantum Materials, School of Physics, Peking University, Haidian District, Beijing 100871, China\\
$^{2}$Collaborative Innovation Center of Quantum Matter, Beijing 100871, China\\
$^{3}$Wuhan National High Magnetic Field Center, Huazhong University of Science and Technology, Wuhan 430074, China\\
}

\begin{abstract}
We report large magnetoresistance (MR) at low temperatures in single-crystalline nonmagnetic compounds TaAs$_2$ and NbAs$_2$.
Both compounds exhibit parabolic-field-dependent MR larger than $5\times10^3$ in a magnetic field of 9 Tesla at 2 K.
The MR starts to deviate from parabolic dependence above 10 T and intends to be saturated in 45 T for TaAs$_2$ at 4.2 K.
The Hall resistance measurements and band structural calculations reveal their compensated semimetal characteristics.
The large MR at low temperatures is ascribed to a resonance effect of the balanced electrons and holes with large mobilities.
We also discuss the relation of the MR and samples' quality for TaAs$_2$ and other semimetals.
We found that the magnitudes of MR are strongly dependent on the samples' quality for different compounds.
\end{abstract}

\pacs{74.70.Xa, 
75.30.Gw, 
75.30.Ds 
}

\maketitle

\section{Introduction}

Magnetoresistance (MR) is defined as the change of material's resistance in magnetic fields ($\mathrm{MR} \equiv \Delta \rho /\rho _{H=0}$).
Recent studies on the single-crystalline semimetals including WTe$_2$\cite{WTe2Cava, WTe2PRLforMobility}, Cd$_3$As$_2$\cite{Cd3As2OngNatMaterials}, NbSb$_2$\cite{NbSb2SciRep}, and MPn (M = Ta and Nb, Pn = P and As)\cite{TAWeylJiagroup, NbAsmobility_Germann, NbPGermann, TaP&NbPJiaGroup, TAothergroup, NbPsecondMR, NbAs_repeated_work, repeated_TA_other_group} have revealed large MR varying from $1\times10^3$ to $2\times10^4$ in a magnetic field less than 10 Tesla (T) at low temperatures.
Other metallic compounds such as PtSn$_4$\cite{PtSn4Canfield} and PdCoO$_2$\cite{PdCoO2PRL} with much higher carriers' density were reported to have large MR being $5.0\times10^3$ and $3.5\times10^2$ in a magnetic field of 14 T around 2 K, respectively.
Compared with the MR for the prototype semimetal element bismuth which was reported to be as large as $1.6\times 10^7$ at 4.2 K in 5 T\cite{Bismuthlowtempnum1}, the MR for these non-magnetic compounds is at least two orders of magnitude less.
Nevertheless, the MR for these compounds was described as ``large'', because the values are several orders of magnitude larger than that for normal metals such as copper \cite{mrinmetals}.

Several mechanisms have been proposed to explain the large MR in non-magnetic semimetals\cite{OxfpreofTransporttheory,Ag2SeNatureFirest}.
A classical two-band model predicts large parabolic-field-dependent MR in a compensated semimetal.
A small difference of the electrons and holes densities ($n$ and $p$) will cause the MR eventually saturated to a field-independent value\cite{WTe2Cava, OxfpreofTransporttheory, MetaltoInsulatorPRL}.
Most of the compounds with large MR show a power law of field dependence very close to parabolic ($\mathrm{MR}\propto H^2$).
The resonance of electrons and holes was believed to occur in bismuth and graphite, while their MR was observed to be saturated in an intense magnetic field.
On the other hand, the MR for WTe$_2$ and NbSb$_2$ are not saturated in a magnetic field as high as 60 T and 32 T, respectively.

Some semimetals with large MR show the features that cannot be simply understood as a resonance of the electron-hole balance.
The large MR for metallic PtSn$_4$ was reported as the result of ultrahigh $\omega _c\tau$ \cite{PtSn4Canfield}.
The linear field dependence of the MR in Cd$_3$As$_2$\cite{Cd3As2OngNatMaterials}, Ag$_2$Se\cite{Ag2Sesinglecryjiagroup} and the recently reported Weyl semimetal TaAs family \cite{TAWeylJiagroup, NbPGermann, TaP&NbPJiaGroup, NAFirstpaper} is not expected for resonant semimetals.
The linear field dependent MR was understood as a quantum effect near the crossing point of the conduction and valence bands which have linear energy dispersion when the magnetic field is beyond the quantum limit \cite{Quantummagnetoresistance, Abrikosov2000Quantum}.
A classical effect of the spacial mobility fluctuation causes linear MR as well, which is more significant in polycrystalline samples \cite{parish2003non}.
Based on the studies of the large MR in these non-magnetic compounds, the mechanisms seem to be different from system to system.
Exploring the semimetals which show large MR but crystallizing in different structures will help to understand this complexity.
Unfortunately very limited semimetal compounds were reported to enact the forms of high-quality single crystals.

TaAs$_2$ and NbAs$_2$ belong to a group of transition metal dipnictides MPn$_2$ (M = V, Nb, Ta, Cr, Mo and W,  Pn = P, As and Sb) which crystallize in an OsGe$_2$-type structure\cite{NbAs2oldeststructureinNature, DetermineCrystruNbAs2}.
This structure has one crystallographic M site and two Pn sites in one unit cell (Fig. \ref{structure}~a).
Two Pn sites are quite distinct in their coordinations: one site is isolated from each other while the other site forms separated Pn-Pn dimers.
The As-As bond lengths of the dimers in TaAs$_2$ and NbAs$_2$ are 2.42 \AA~and 2.44 \AA ~respectively\cite{DetermineCrystruNbAs2, HintforTaAs2crystalgrowth}, which are very close to the covalent radii 2.43 \AA~ of the As-As dimer in tetramethyldiarsane\cite{dimerAs-As}.
Each tantalum and niobium atom is surrounded by six As atoms forming a trigonal prism with two additional As atoms outside the rectangular faces.
The prisms are stacked along the crystallographic $b$ direction through their trigonal faces to form the structure.
A naive expectation for the electronic states of NbAs$_2$ and TaAs$_2$ is based on the simple Zintl concept\cite{Zintlconceptview, BandcalwrongNA2Struc}:
a balanced state for valence electrons of (Nb/Ta)$_2^{5+}$(As-As)$^{4-}$As$_2^{6-}$ indicates an insulating state with a gap between the band constructed by the bonding orbitals and antibonding orbitals of the As-As dimers.
However previous band-structure calculations\cite{BandcalwrongNA2Struc} suggested that NbAs$_2$ and TaAs$_2$ are semimetals in stead of semiconductors.
Although the Nb-Nb or Ta-Ta bonds with the bond-length close to 3 \AA~are not considered in this oversimple Zintl concept, this rough estimation predicts the destitute density of the states (DOS) for TaAs$_2$ and NbAs$_2$ near the Fermi level.

Physical properties for the MPn$_2$ family have not been extensively studied.
MoAs$_2$ and WAs$_2$ were reported to be superconductors with the $T_c$ less than 1 K\cite{BandcalwrongNA2Struc}.
The semimetal behaviors and large, parabolic field-dependent MR at low temperatures in NbSb$_2$ were reported very recently\cite{NbSb2SciRep}.
The Sb atoms in NbSb$_2$ form the dimers with the Sb-Sb bond-length of 2.71 \AA ~as well.
The band structure calculations for NbSb$_2$ show compensated, small electron and hole pockets near the Fermi level\cite{NbSb2SciRep}.
A resonance of the electron-hole balance in NbSb$_2$ was suggested to be the reason of the large MR at low temperatures\cite{NbSb2SciRep}.

Due to highly volatile and toxic arsenic, single crystals of the transition metal diarsenides cannot be easily obtained via a regular flux growth as described in Ref.\cite{NbSb2SciRep, WTe2PRLforMobility}.
No physical properties have ever been reported for the single-crystalline NbAs$_2$ and TaAs$_2$.
In this paper, we show that large single crystals of NbAs$_2$ and TaAs$_2$ can be grown via a chemical vapor transport (CVT) process.
We found that these two compounds are compensated semimetals with large, parabolic-field-dependent MR at low temperatures.
Based on the analysis of the electrical properties and band structure calculations, we believe that their large MR are due to a resonance of the electron-hole balance.
After an analysis of the semimetals with large MR in previous reports, we found that the values of MR are strongly dependent on the samples' quality.
We believe that most of the parabolic-field-dependent large MR for compensated semimetals can be explained as a result of electron-hole balance.

\section{Experiment}

\begin{figure}
\begin{center}
\includegraphics[width=6in]{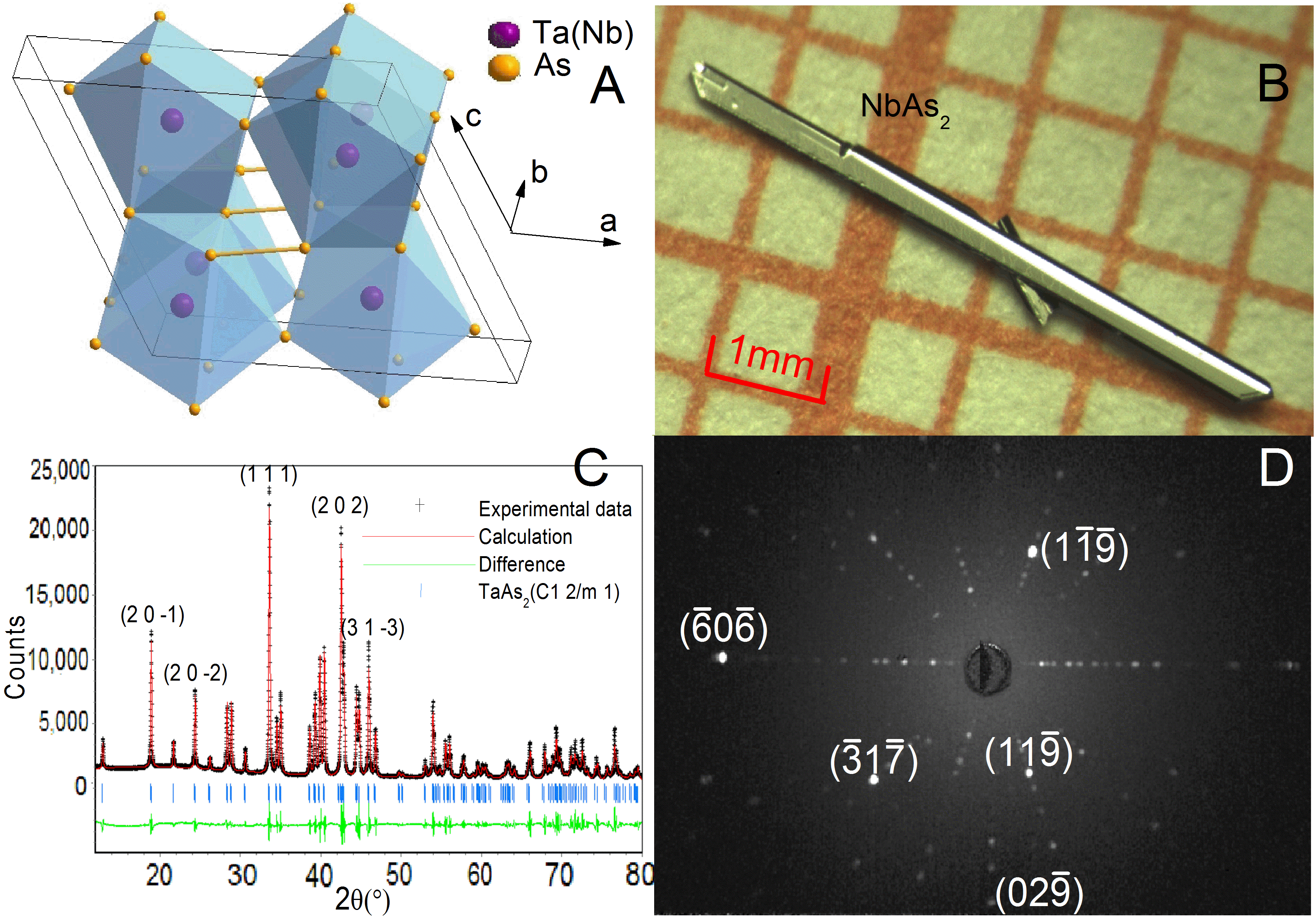}
\caption{Panel a: Crystal structure for Ta(Nb)As$_2$.
Panel b: Photo of a single crystal of NbAs$_2$ on millimeter-grid paper.
Panel c: Powder XRD spectrum for TaAs$_2$. The refinement results are close to those in Ref.~\cite{DetermineCrystruNbAs2, HintforTaAs2crystalgrowth}.
Panel d: Laue back-scattering diffraction pattern for a single crystal of NbAs$_2$.
}
\label{structure}

\end{center}
\end{figure}

Single crystals of TaAs$_2$ and NbAs$_2$ were found as by-products when we tried to grow the single crystals of TaAs and NbAs via a CVT reaction method.
Then we optimized the growth conditions in the previous report \cite{HintforTaAs2crystalgrowth} for high-quality, large single crystals.
Polycrystalline TaAs$_2$ and NbAs$_2$ was prepared by heating the stoichiometric amounts of elemental Ta/Nb and high-purity arsenic in evacuated silica ampules at 750~$^\circ$C for 2 days.
Some redundant arsenic grains were always observed in the ampules after the synthesis.
This phenomenon is consistent with the observation of the vacancy of arsenic atoms in TaAs$_2$ and NbAs$_2$ as described in Ref.~\cite{DetermineCrystruNbAs2, HintforTaAs2crystalgrowth}.
The crystals of TaAs$_2$ were grown in a temperature gradient from 850~$^\circ$C (source) to 750~$^\circ$C (sink) for 7 days with the concentration of the agent TaBr$_5$ being 2.7~mg/cm$^3$, while the crystals of NbAs$_2$ were grown in a temperature gradient from 775~$^\circ$C (source) to 700~$^\circ$C (sink) for 7 days with the concentration of the agent NbI$_5$ being 2.7~mg/cm$^3$.
Blade-like crystals yielded in these conditions have the length about 3-7 mm and the width about 0.6 mm (Fig.~\ref{structure}~b) while the residue resistivity ratio ($\mathrm{RRR} = R_{300K}/R_{2K}$) is larger than 100 in transport measurements.
CVT growth from the polycrystalline TaAs and NbAs as the sources in stead of TaAs$_2$ and NbAs$_2$ and TeCl$_4$ (1.5~mg/cm$^3$) or TeI$_4$ (1.9~mg/cm$^3$) as the agent also yielded needle-like single crystals of TaAs$_2$ and NbAs$_2$ but their $\mathrm{RRR}$ is much smaller.

Powder X-ray diffraction (XRD) measurements were performed in a Rigaku MiniFlex 600 diffractometer.
The refinements by using the Rietica Rietveld program confirm that the compounds crystallize in the structures of the space group $C2/m$, same as  previously reported \cite{DetermineCrystruNbAs2}.
The refined parameters are $a=9.3372(2)$ \AA , $b=3.3847(1)$ \AA, $c=7.7711(2)$ \AA~and $\beta=119.638(1)^{\circ }$ for TaAs$_2$ and $a=9.4017(2)$ \AA,  $b=3.3947(1)$ \AA, $c=7.8212(2)$ \AA, and $\beta=119.531(1)^{\circ}$ for NbAs$_2$ (Fig.~\ref{structure}~c).
Laue XRD measurements (Fig.~\ref{structure}~d) on the single crystals were collected in a Photonic Science PSL-Laue Back-scattering System.
The pattern shows that the long direction of the crystals is the $b$ direction while the largest flat plane in blade-like crystals is perpendicular to the (001) direction (Fig.~\ref{structure}~d).
The resistivity and the Hall resistance were measured with a standard 4-probe ac apparatus\cite{thermoelectricity} in which the current is parallel to the $b$-axis and the magnetic field is parallel to the (001) direction.
The resistance and Hall data have been symmetrized to remove the effect of the misalignment of voltage wires ($\rho_{xx}(H)=\frac{\rho_{xx}(H)+\rho_{xx}(-H)}{2}$, $\rho_{yx}(H)=\frac{\rho_{yx}(H)-\rho_{yx}(-H)}{2}$).
All the physical property characterizations were performed in a Quantum Design Physical Property measurement system (PPMS-9) over the temperature range from 1.9 K to 300 K and in the magnetic fields up to 9 T.

Band structure calculations were performed by using the WIEN2K package\cite{WIEN2K} based on the full potential linearized augmented plane wave (LAPW) method\cite{ForWIEN2K} within local density approximation (LDA)\cite{AccforWIEN2K}.
The lattice constants were chosen from Ref.~\cite{DetermineCrystruNbAs2}.
The values of the  smallest atomic sphere radii in the unit cell (R$_{mt}$) were set to 2.48 Bohr and 2.22 Bohr for Ta and As in TaAs$_2$ (2.43 Bohr and 2.20 Bohr for Nb and As in NbAs$_2$) by considering the difference of their nearest neighbor distances.
The general gradient approximation (GGA) proposed by Perdew et al\cite{GRGforWIEN2K} was used for exchange-correlation potential.
A mesh of $21\times21\times21$ $\vec{k}-$points was selected for Brillouin zone integration with R$_{mt}$ $\ast$ K$_{max}$ = 7.0 and Fermi surface plot, where K$_{max}$ is the magnitude of the largest reciprocal lattice vector (K).


\section{Results}

\begin{figure}
\begin{center}
\includegraphics[width=6in]{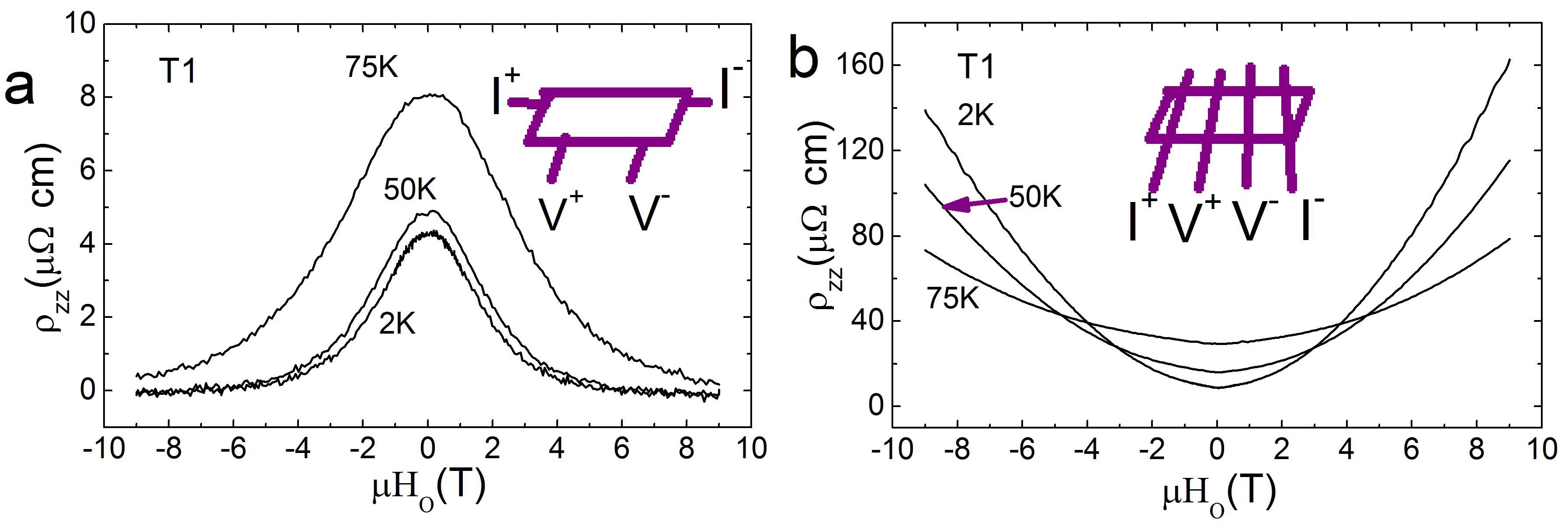}
\caption{Panel a: Measured `longitudinal MR' for the sample T1 of TaAs$_2$ when the contacts are NOT fully crossing the sample as illustrated in the sketch. Panel b: Measured longitudinal MR for the same sample when the contacts are fully crossing the sample as illustrated in the sketch.
}
\label{negative}
\end{center}
\end{figure}

Before starting to describe the transport properties for TaAs$_2$ and NbAs$_2$, we present a misleading measurement result due to an improper experimental setup.
Figure \ref{negative} a and b show the measured resistivity for a same sample T1 with different contact setups as the insets, respectively, when the magnetic field is parallel with the current (H $\parallel $i).
The contacts attached the edge of the sample in the first setup which is commonly employed for measuring Hall signal and resistance at the same time.
As shown in Fig.~\ref{negative} a, a very large negative ``longitudinal MR'' occurs in the measurements for this setup.
However when the contacts fully crossed the wideness in the second setup, the MR becomes a regular, small, positive profile (Fig.~\ref{negative} b).
We believe that the negative ``longitudinal MR'' is due to a magnetic-field-induced current jetting effect which is commonly observed in high mobility samples \cite{CurjettingforAg2Se, LinearMRInSilicon, mrinmetals}.
The current jetting effect must be seriously considered for measuring the longitudinal MR, especially for exploring the chiral anomaly in recently discovered Weyl semimetals\cite{TAWeylJiagroup}\cite{TaP&NbPJiaGroup}.

\begin{figure}
\begin{center}
\includegraphics[width=6in]{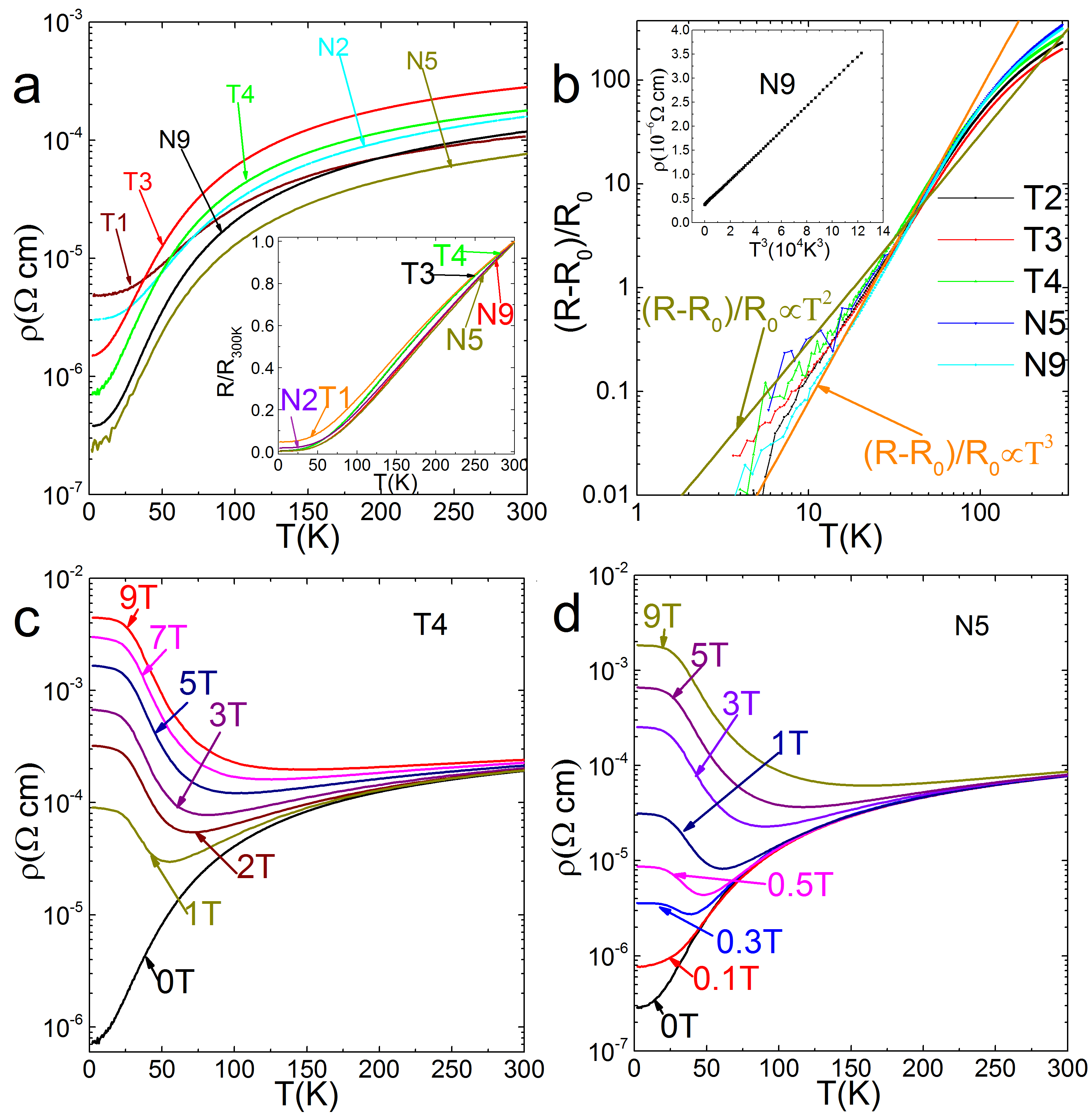}
\caption{Panel a: Temperature dependent resistivity for different samples of TaAs$_2$ and NbAs$_2$ in a semi-logarithmic plot (N for NbAs$_2$ and T for TaAs$_2$). Inset: $R(T)/R_{300K}$ versus T.
Panel b: A double-logarithmic plot for the resistance and temperature. Inset: Resistivity versus T$^3$ for N9.
Panel c and d: Temperature dependent transversal resistivity for the samples T4 and N5 in different magnetic fields, respectively.}
\label{RT}
\end{center}
\end{figure}

Figure \ref{RT} a shows a metallic profile for the temperature dependent resistivity of several representative samples for TaAs$_2$ and NbAs$_2$ in zero magnetic field.
With a similar profile of linear-temperature-dependent resistivity above 100 K, the samples T3, T4, N5 and N9 which were grown by using TaBr$_5$ or NbI$_5$ as transfer agents have much smaller residue resistivity at 2 K than those of the samples T1 and N2 which were grown by using I$_2$ as agent.
Below 50 K, the samples with large $\mathrm{RRR}$ exhibit the R-T profiles with the relation of $R(T) = R_0 + aT^n$ for both compounds.
As shown in Fig. \ref{RT} b, $n$ equals 3.038(4) for N9, while it varies from 2.7 to 2.9 for the other four samples.
This power law dependence is distinct from a quadratic temperature dependent behavior for Fermi liquid.
A T$^3$ dependent resistance is observed in transition metal elements as well, which is ascribed as a scattering effect of the conduction electrons into vacant states of $d$-band \cite{book:1130301}.

When a magnetic field is applied to the samples T4 and N5 with large $\mathrm{RRR}$, their resistivity at low temperatures significantly increases while the increase at high temperatures is very limited (Fig. \ref{RT} c and d).
The large difference of the MR at high and low temperatures leads $\rho (T)$ in a magnetic field showing a ``transition'' from  a metallic profile at high temperatures to a semiconducting profile at low temperatures.
This field-induced transition-like behavior is similar to the previously reported behavior in other semimetals \cite{WTe2Cava, NbSb2SciRep, PtSn4Canfield, TAWeylJiagroup, TaP&NbPJiaGroup, NbPGermann, TaP&NbPJiaGroup, MetaltoInsulatorPRL}.

\begin{figure}
\begin{center}
\includegraphics[width=6in]{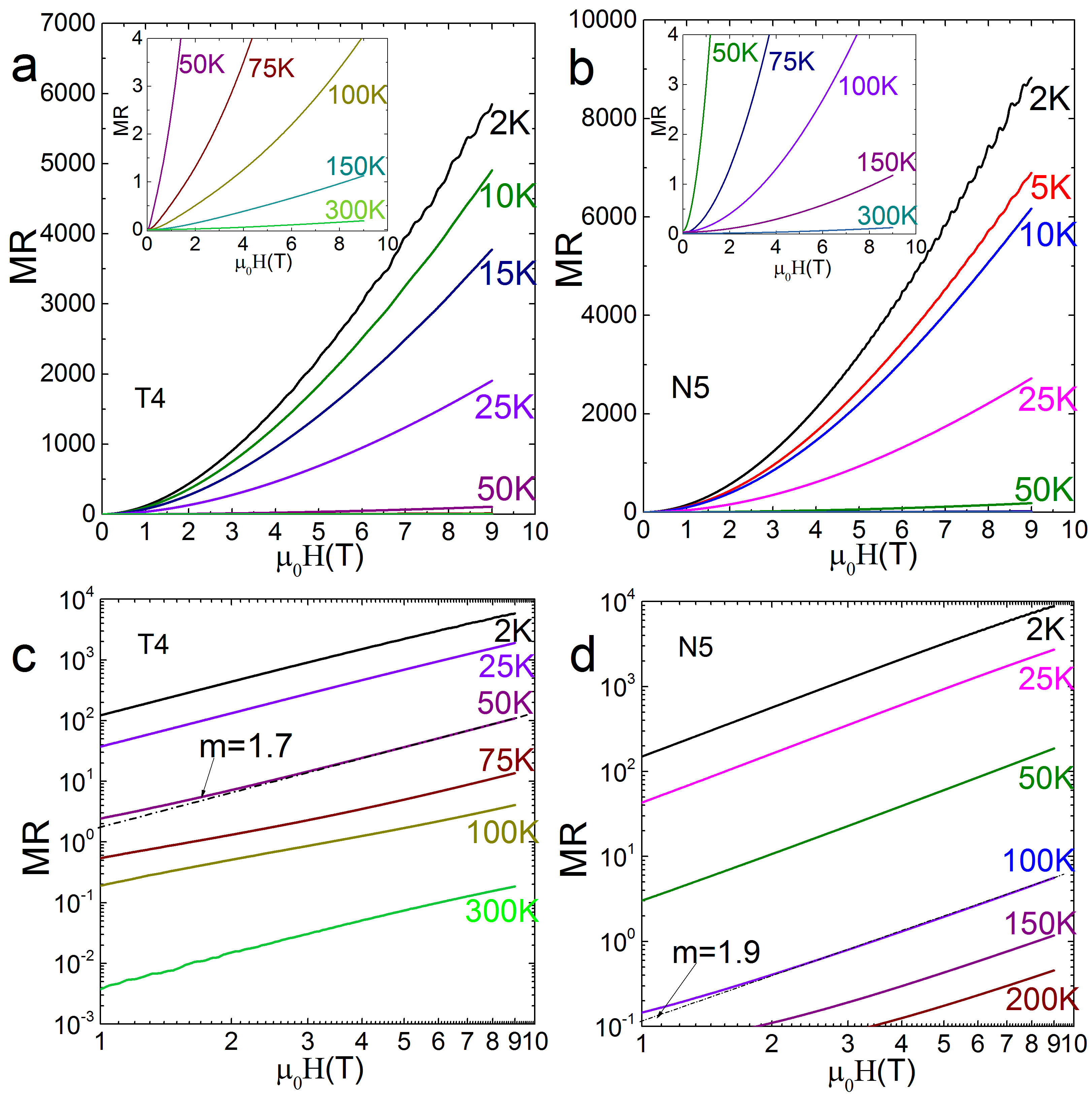}
\caption{Panel a and b: The MR versus magnetic field at different temperatures for the sample T4 and N5, respectively. Insets: The MR at high temperatures. Panel c and d: the MR in double logarithmic plots for T4 and N5, respectively. MR $\propto H^m$ at low temperatures where $m=1.9$ and $1.7$ for N5 and T4, respectively.
}
\label{RH}
\end{center}
\end{figure}

Figure \ref{RH} a and b show that the MR values decay three orders of magnitude from 2 K to 100 K for T4 and N5.
Below 5 K, the MR shows clear Shubnikov-de Hass (SdH) oscillations on the background of a parabolic profile.
Although the MR changes significantly at different temperatures, it remains a clear power-law dependence with $MR\propto H^m$ where $m=1.7$ and $1.9$ for T4 and N5, respectively (Fig. \ref{RH} c and d).
This power law is very close to a parabolic field dependence predicted by the two-band theory for identical compensated hole and electron densities\cite{WTe2Cava, OxfpreofTransporttheory, MetaltoInsulatorPRL}.
The MR persists this power law dependence but it is only 0.13 and 0.19 for N5 and T4 in 9 T at room temperature (300K), respectively.
We notice that some semimetals including bismuth and graphite \cite{BismuthvalleyNatPhy, MR_in_graphite_useful_1, Ag2Sesinglecryjiagroup, Cd3As2OngNatMaterials, TAWeylJiagroup, NAFirstpaper, NbPGermann, TaP&NbPJiaGroup} show their MR one to two orders of magnitude larger than those of TaAs$_2$ and NbAs$_2$ at room temperature.
On the other hand, the MR values for T4 and N5 at room temperature are close to the reported values for WTe$_2$\cite{WTe2Cava}, NbSb$_2$\cite{NbSb2SciRep}, PdCoO$_2$\cite{PdCoO2PRL} and PtSn$_4$\cite{PtSn4Canfield}.

\begin{figure}
\begin{center}
\includegraphics[width=6in]{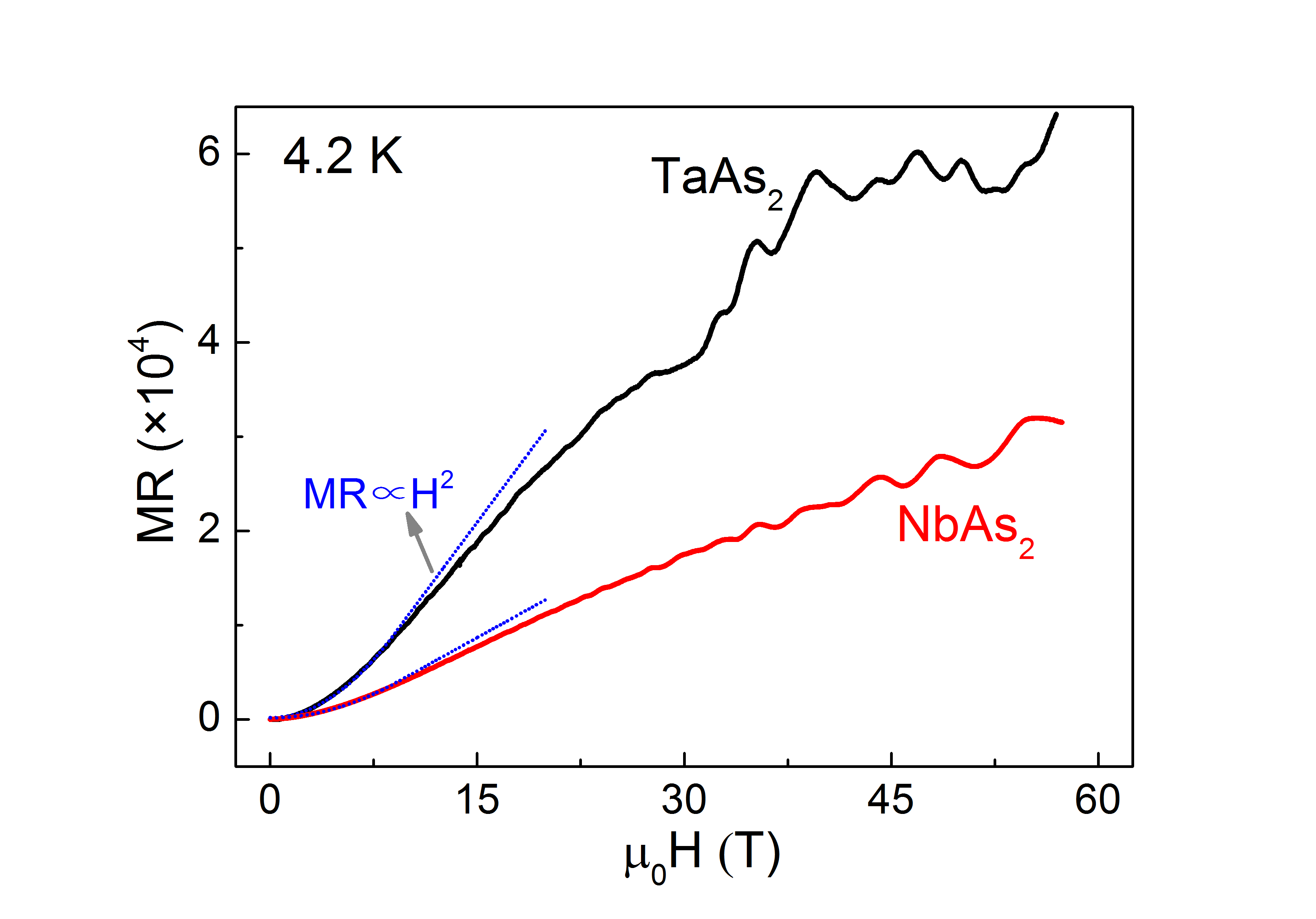}
\caption{MR for TaAs$_2$ and NbAs$_2$ in a pulsed magnetic field at 4.2 K.
}
\label{highfield}
\end{center}
\end{figure}

In order to better understand the MR for TaAs$_2$ and NbAs$_2$, two samples were measured in a pulsed magnetic field as high as 55 T in National High Magnetic Lab in WuHan, China.
Figure \ref{highfield} shows the preliminary results for both compounds.
The MR deviates from the quadratic field-dependent power law above 15 T.
The MR for TaAs$_2$ shows strong, complicated SdH oscillations above 30 T and intends to be saturated at 45 T.
Detail analysis on the SdH oscillations in high magnetic field will be presented in the future.

\begin{figure}
\begin{center}
\includegraphics[width=6in]{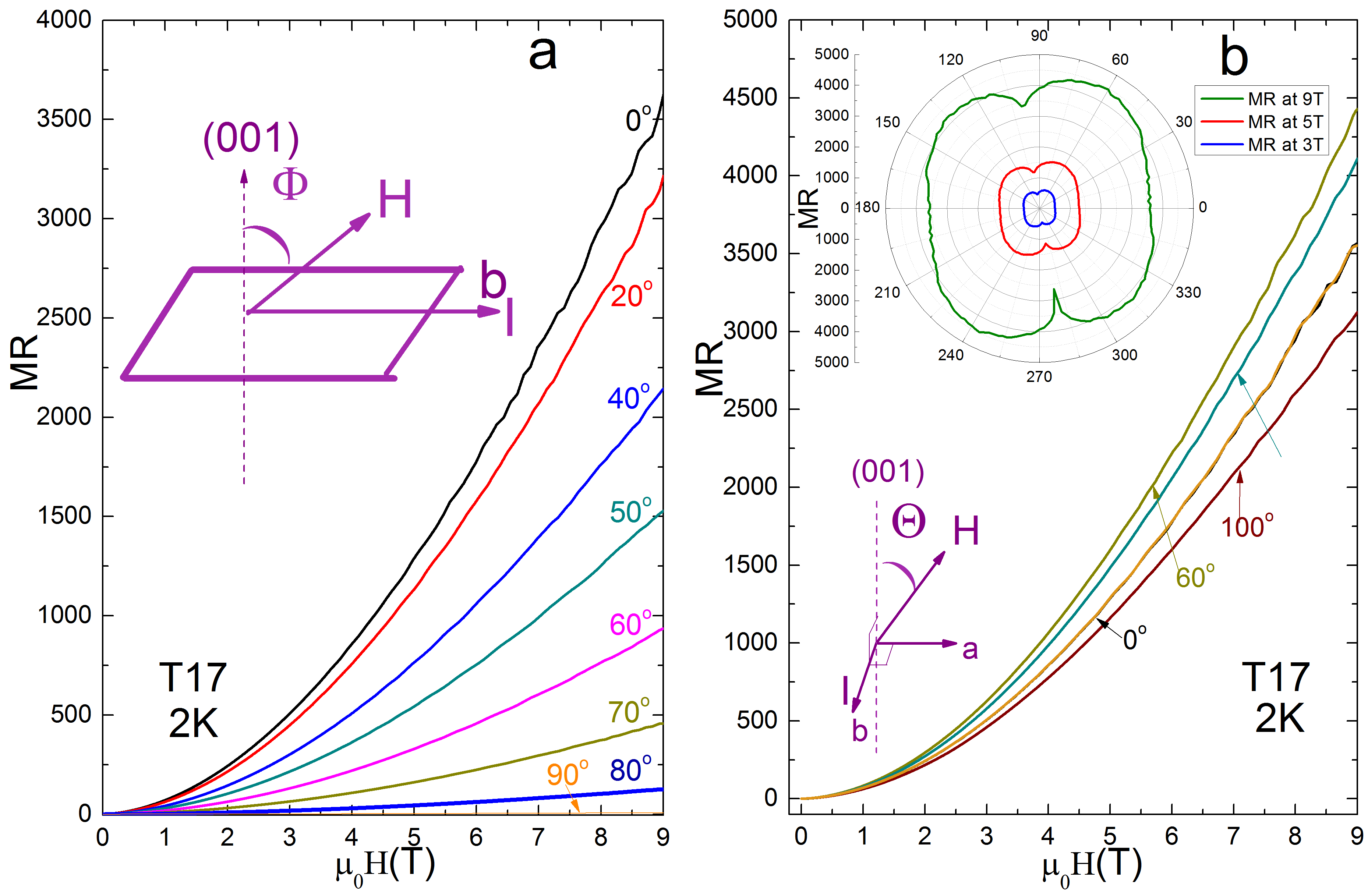}
\caption{Panel a: MR versus H at 2 K when H is tilted from (001) direction to $b$ direction.Inset: the sketch of the experimental set-up.
Panel b:  Transversal MR versus H at 2 K when H is rotated in the plane perpendicular to $b$ direction.
Upper inset: Angular dependent transversal MR at 2 K for H fixed in 3, 5 and 9 T.
Lower inset: the sketch of the transversal MR experimental set-up.
The angular dependence of the MR for NbAs$_2$ is similar to that for TaAs$_2$.
}
\label{rotation}
\end{center}
\end{figure}

Figure \ref{rotation} shows the MR when the magnetic field is tilted in different directions while the current is always along $b$ direction.
In order to minimize the current jetting effect, the sample T17 was prepared to be long, thin bars with four contacts fully crossed their wideness in this measurement.
When the field direction is tilted from that parallel to the (001) direction to $b$ direction, the MR drops rapidly (Fig. \ref{rotation} a).
Such large difference between the longitudinal and transversal MR is also observed in WTe$_2$\cite{WTe2Cava} and NbSb$_2$\cite{NbSb2SciRep}.
On the other hand, the transversal MR changes small when the magnetic field is rotated in the plane perpendicular to $b$ direction.
Figure \ref{rotation} b shows that the minimal of the transversal MR occur when the field is close to the $a$ direction ($\Theta = 280^{\circ } $ and $100^{\circ } $).
The polar plot for the MR in the inset of Fig. \ref{rotation} b shows a two-fold-rotation symmetric pattern, which is consistent with the monoclinic  crystal structure.

\begin{figure}
\begin{center}
\includegraphics[width=6in]{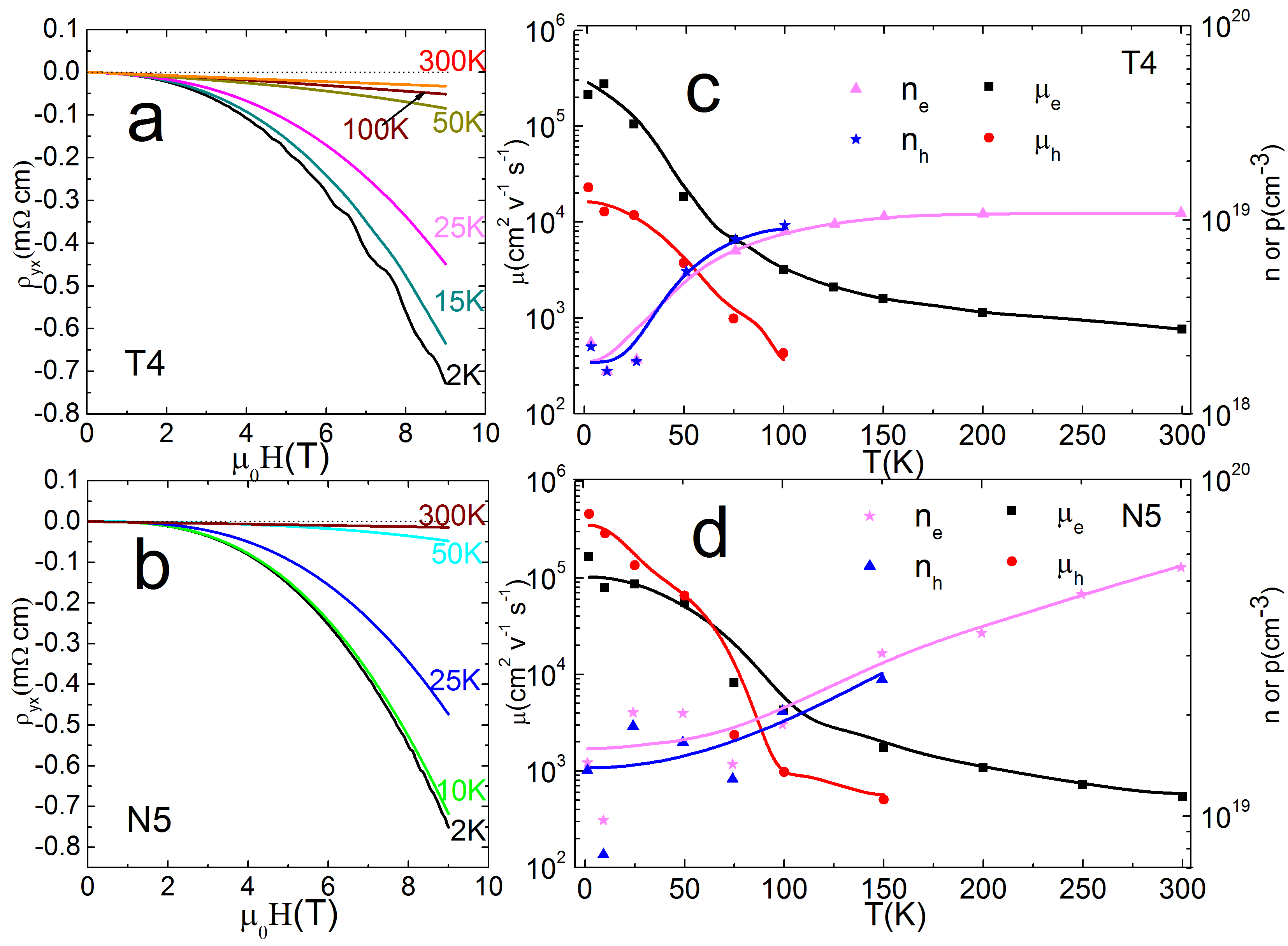}
\caption{Panel a and b: Field dependent Hall resistivity at different temperatures for the samples T4 and N5, respectively. Panel c and d: The fitting results for the electron and hole densities ($n$ and $p$), and their mobilities ($\mu _e$ and $\mu _h$) at different temperatures for T4 and N5, respectively. No hole information above 150 K is obtained in this measurement. The lines are guided by eyes.
}
\label{Hall}
\end{center}
\end{figure}

Hall measurements provide informations about the carriers for TaAs$_2$ and NbAs$_2$.
Figure \ref{Hall} a and b show that the Hall Resistivity ($\rho _{yx} $) is negative and linear dependent to the  magnetic field when $T \geq $ 125 K and 200 K for T4 and N5, respectively.
At low temperatures, $\rho _{yx} $ apparently deviates from the linear-field dependence, which indicates two types of carriers.
We fitted the Hall conductivity ($\sigma_{xy}$) by a two-carrier model derived from the two-band theory\cite{Hurd1972Hall_effect_in_metal},
\begin{equation}
\begin{split}
\sigma_{xy}=[n_h\mu_h^2\frac{1}{1+(\mu_hH)^2}-n_e\mu_e^2\frac{1}{1+(\mu_eH)^2}]eH
\end{split}
\end{equation}
where $\sigma_{xy}=\frac{\rho_{yx}}{\rho_{yx}^2+\rho_{xx}^2}$, n$_e$ (n$_h$) and $\mu_e$ ($\mu_h$) denote the carrier concentrations and mobilities for electrons (holes), respectively.
The fitting results  show close $n$ and $p$ at low temperatures (Fig. \ref{Hall} c and d).
At high temperature range, the hole contributions to the Hall signals are negligible due to its rapidly decreased mobility, and the data can be well fitted by an one-band model.
The mobilities at 2 K are about $1\times10^5$ cm$^2$V$^{-1}$s$^{-1}$ for electrons and hole for both TaAs$_2$ and NbAs$_2$.
At room temperature, $\mu_e$ drops to the order of $5\times10^2$ cm$^2$V$^{-1}$s$^{-1}$ for both samples.
The Hall measurements revealed that TaAs$_2$ and NbAs$_2$ are compensated semimetals with large mobilities of electrons and holes at low temperatures.

\begin{figure}
\begin{center}
\includegraphics[width=6in]{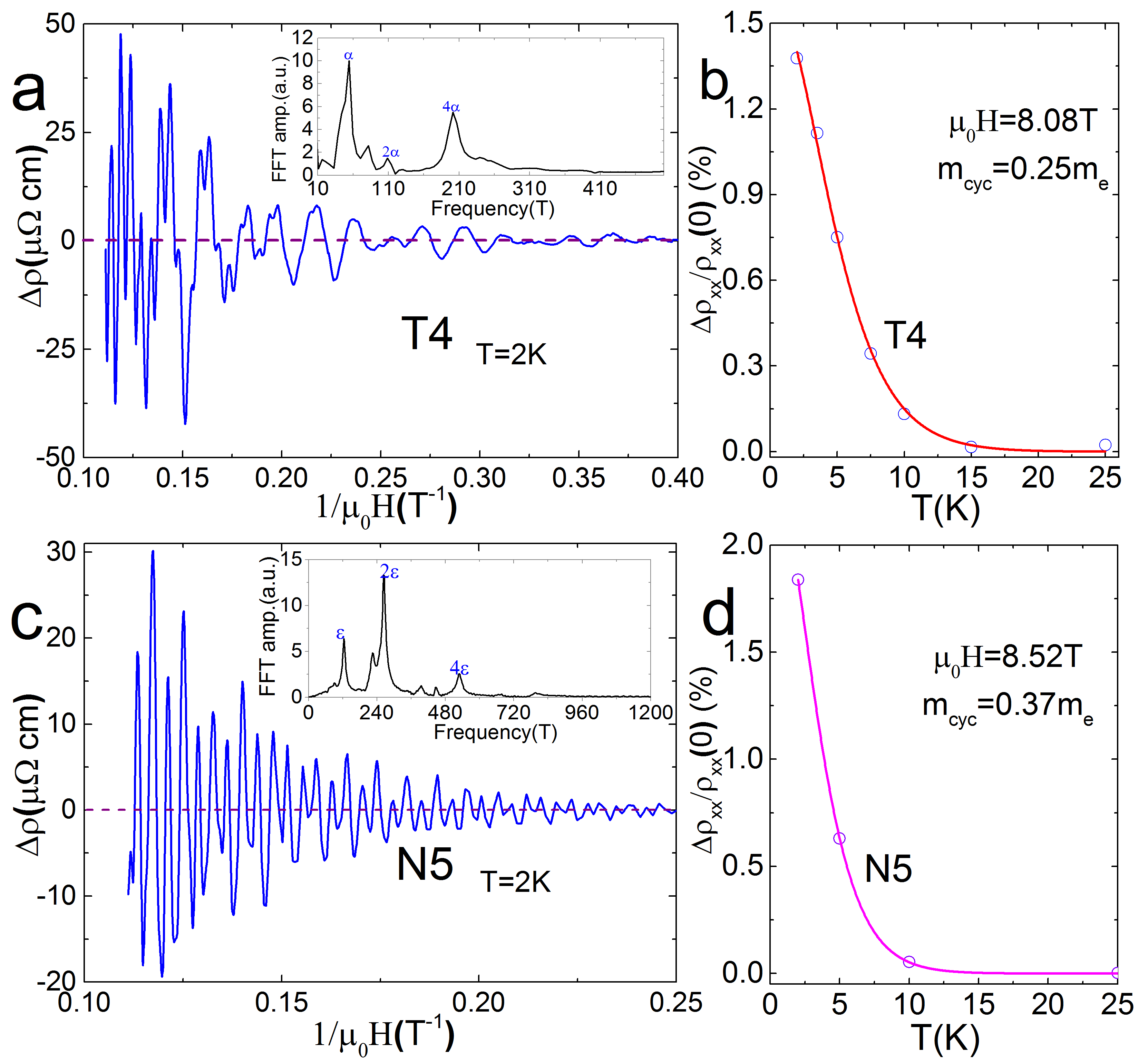}
\caption{Panel a and c: Oscillatory part of the resistivity versus $1/\mu _0H$ for the samples T4 and N5 at 2 K, respectively. Insets: FFT spectra for T4 and N5. Panel b and d: The fitting of the amplitude of the SdH oscillations with respect to temperature yields the cyclotron electron mass for the main frequencies for T4 and N5, respectively.
}
\label{SdH}
\end{center}
\end{figure}

The SdH oscillations in the field-dependent resistivity for TaAs$_2$ and NbAs$_2$ were extracted via subtracting the non-oscillation background at low temperatures (Fig. \ref{SdH}).
The SdH oscillations with respect to $1/H$ form complicated patterns which obviously come from multiple frequencies.
Fast Fourier Transform (FFT) analysis reveals that the samples T4 and N5 have a main frequency about $\alpha$ = 50 T and $\varepsilon$ = 120 T, respectively, while double and quadruple harmonic frequencies of the main were observed as well.
The existence of multi-harmonic frequencies indicate a long transport lifetime for both compounds in accordance with their high mobilities.
In order to obtain the cyclotron mass for the main frequencies $\alpha$ and $\varepsilon$, we fitted $\Delta\rho_{xx}/\rho_{xx}$ maximum amplitude at $\mu _0H = 8.08$ T and $8.52$ T vs. temperature for T4 and N5, respectively (Fig. \ref{SdH} b and d).
Using the fitting formula\cite{BerryphaseScience}
\begin{equation}
\begin{split}
\frac{\Delta\rho_{xx}}{\rho_{xx}}\propto\frac{2\pi^2k_BT/\hbar\omega_c}{sinh(2\pi^2k_BT/\hbar\omega_c)}
\end{split}
\end{equation}
where $k_B$ is Boltzmann constant, cyclotron frequency $\omega_c$ = $eB/m_{cyc}$, and $m_{cyc}$ is cyclotron mass,
we got $m_{cyc}$ equals to 0.25 m$_e$ and 0.37 m$_e$ for T4 and N5, respectively.
Further experiments acompanied by acute band structure calculations are needed to address the pockets related to these SdH oscillations.

\section{Discussion}

\begin{figure}
\begin{center}
\includegraphics[width=6in]{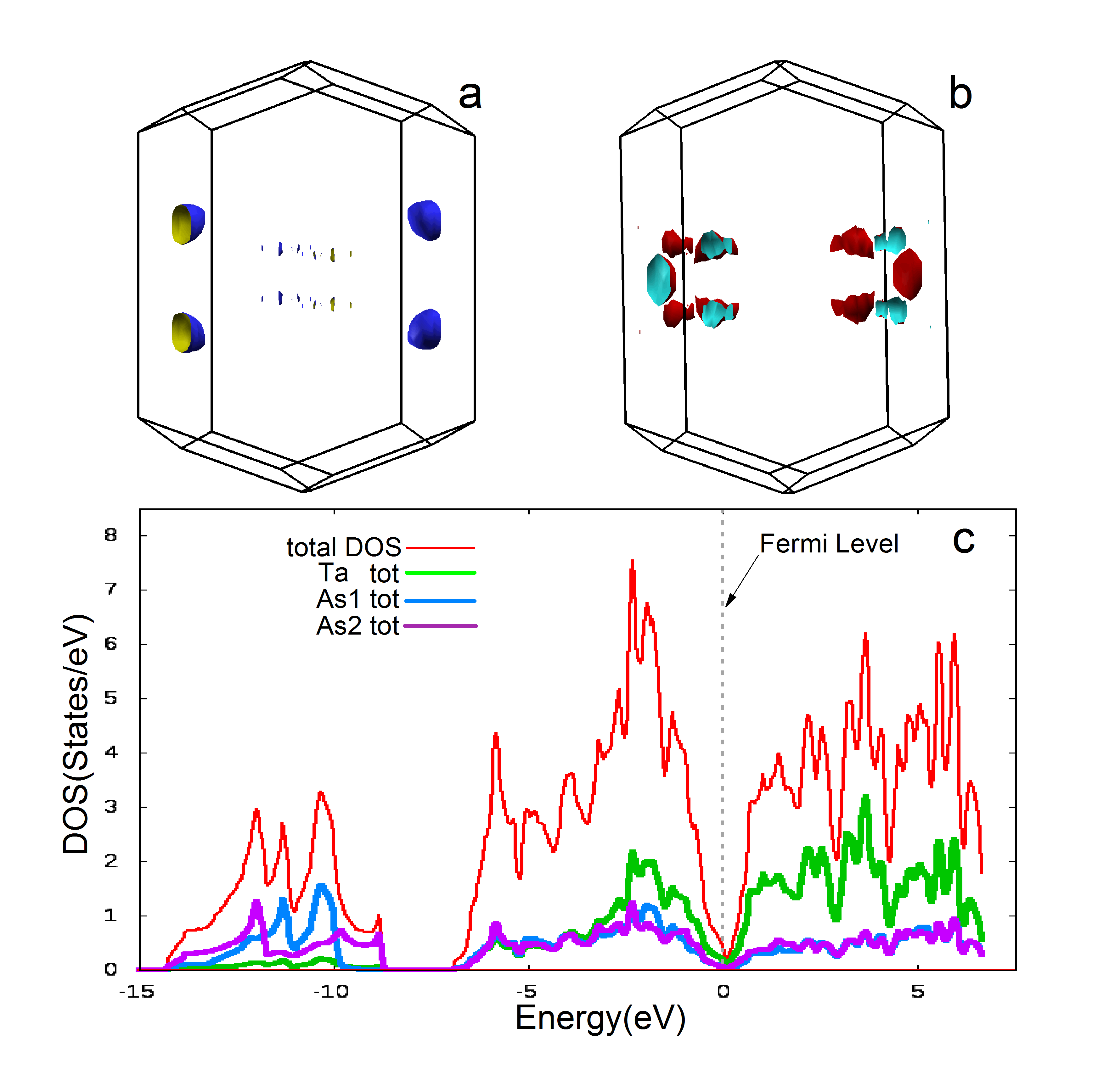}
\caption{Panel a and b: Fermi Surfaces of electron and hole pockets for TaAs$_2$ in its primitive Brillouin zone, respectively.
Panel c: The total and partial DOS for TaAs$_2$.
}
\label{band}
\end{center}
\end{figure}

The calculated band structures for TaAs$_2$ and NbAs$_2$ are consitent with their compensated semimetal characteristics.
Figure \ref{band} a and b show the Fermi surface mapping in which only one valence and conduction bands intersect the Fermi level, respectively.
Small, ellipse-like electron and hole pockets appear at the boundary of the primitive Brillouin zone.
The DOS at Fermi level is very close to the minimum in the spectrum (Fig. \ref{band} c).
The results for TaAs$_2$ and NbAs$_2$ are similar as that in ref. \cite{dimerAs-As} and the band structure for NbSb$_2$ \cite{NbSb2SciRep}.
Since the Fermi surface is complicated with several different electron and hole pockets, the identification of the frequencies for the SdH oscillations with the extremal surface areas needs further elaborations.

\begin{figure}
\begin{center}
\includegraphics[width=6in]{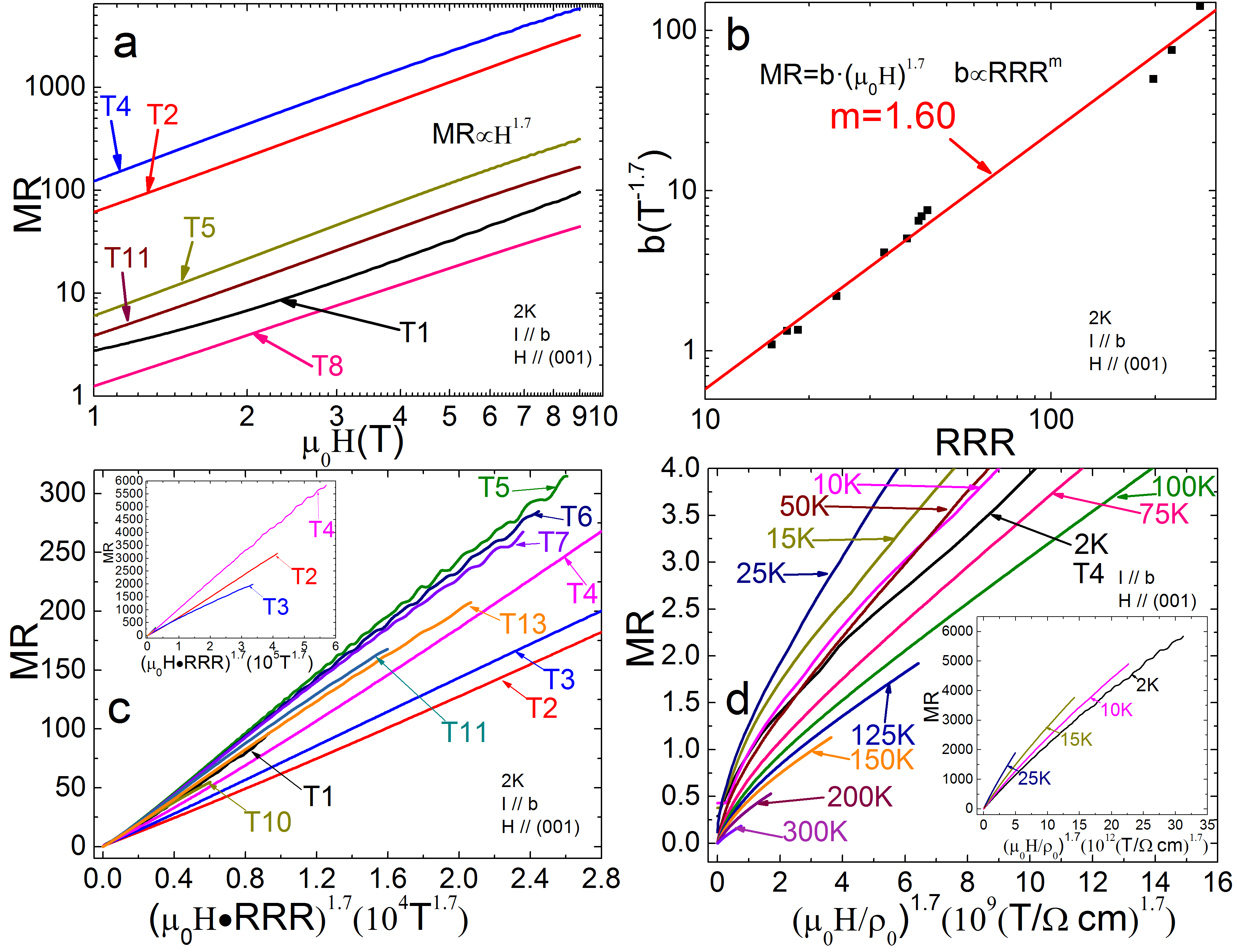}
\caption{Panel a : MR versus magnetic field for different samples of TaAs$_2$ at 2~K in a double-logarithmic plot.
Panel b : Coefficient b (MR$ = bH^{1.7}$) versus RRR for different samples of TaAs$_2$ in a double-logarithmic plot. The red line is the fitting line of the order of $1.6$.
Panel c :Modified Kohler plot for different samples of TaAs$_2$ at 2 K.
Inset: The Kohler plot for the samples T2, T3 and T4 at 2K in different range.
Panel d:  Kohler plot for the sample T4 at different temperatures using $MR=F[H/\rho(0)]\propto {(H/\rho(0))}^{1.7}$.
Inset: The same Kohler plot of Panel d in different range.
}
\label{Kohler}
\end{center}
\end{figure}

We next discuss the sample and temperature dependence for the large MR for TaAs$_2$.
Table \ref{table1} shows that the values of $\mathrm{RRR}$ and MR are strongly dependent on the transfer agents using in the growth.
Althouth the MR changes four orders of magnitude for different samples, it follows the same power law of $\mathrm{MR} = bH^n$, where $n=1.7\pm 0.1$ for all the smaples (Fig. \ref{Kohler}).
On the other hand, the coefficient $b$ seems to follow a power law of $b\propto RRR^m$ where $m=1.6$ for all TaAs$_2$ samples.
This power law relation of the $\mathrm{RRR}$ and $b$ results in the giant difference of the MR for the samples with different quanlities.
The dependence of the MR and $\mathrm{RRR}$ on transfer agents is likely due to the doping effect of the impurity from transport agent, such as tellurium atoms on the As site, or the arsenic deficiency described in Ref.~\cite{HintforTaAs2crystalgrowth}.
Similar significant drop of large MR due to chemical doping has been observed in WTe$_2$ \cite{WTe2doping}.

In order to better understand the electron scattering in diffent samples, the MR at 2 K for different samples is plotted in a modified Kohler plot (Fig.~\ref{Kohler} c).
Considering the large uncertainty of the dimensions for the samples, we chose $(\mu _0\mathrm{RRR})^{1.7}$ instead of $(\mu _0/\rho _0)^{1.7}$ in the plot.
We notice that the curves for the samples with relatively small MR ($<100$) cluster together, while the curves for the three samples of T2, T3 and T4 with large MR ($>1000$) clearly deviate from the cluster.
A plausible explanation is that the impurity induced electron scattering dominates the transport behaviors in a magnetic field for those samples with small RRR.
For high-quality samples such as T2, T3 and T4, the scattering process is strongly dependent on the samples but the orgin is not clear at this point.
A modified Kohler plot for the MR versus  $(\mu _0/\rho _0)^{1.7}$ for the sample T4 at different temperatures is shown in Fig. \ref{Kohler} d.
The curves do not fall into a same region at any temperatures.
Since $\mu _e$ and $\mu _n$ change three orders of magnitude at high and low temperatures, the scattering processes in a magnetic field is plausible to be different at different temperatures.

We list the parameters for the compensated semimetals with large, parobolic-field-dependent MR in Table \ref{table2}.
All the semimetals with large MR have their main-carrier mobilities much larger than $1$m$^2$V$^{-1}$s$^{-1}$.
It seems that the MR for all the semimetals is possitively correlated with the $\mathrm{RRR}$ and mobilities, despite the large difference of the band structures and carrier densities.

\begin{figure}
\begin{center}
\includegraphics[width=6in]{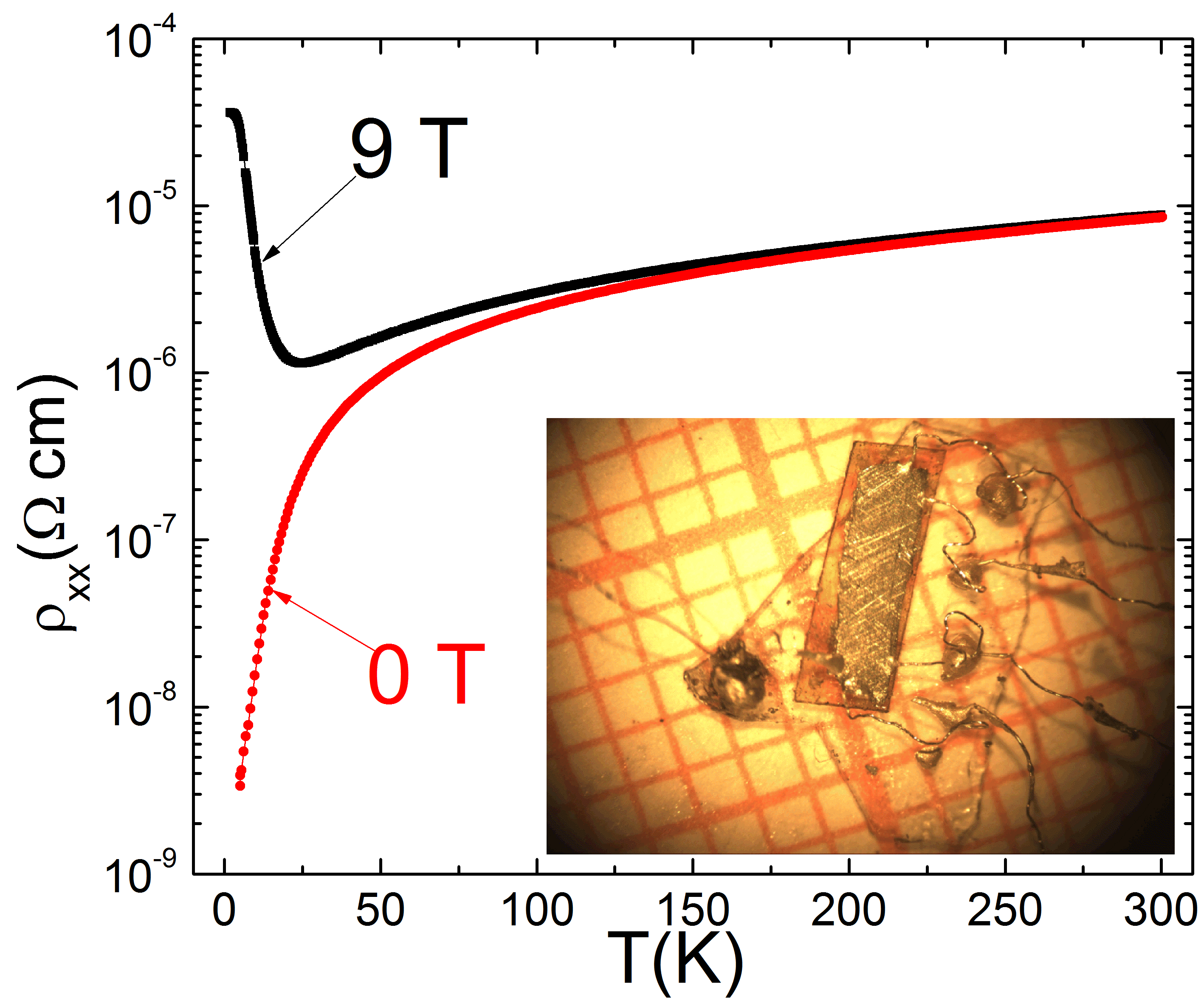}
\caption{Resistivity of a piece of cadmium versus temperature in zero field and 9 Tesla.
Inset : A photo of the polished cadmium bar measured by five-probe technique with millimeter-grid paper background.
}
\label{cadmium}
\end{center}
\end{figure}

By any chance, we measured the MR for a piece of cadmium at 2 K in a magnetic field of 9 T.
This sample was cut and polished from a polycrystalline shot with 99.9997$\%$ purity from Alfa Aeser (see the inset in Fig.~\ref{cadmium}).
Figure ~\ref{cadmium} shows that the R(T) curve in 9 T for polycrystalline cadmium also manifest the ``transition'' from  a metallic profile at high temperatures to a semiconducting profile at low temperatures.
Due to a large current heating effect, the measurement for the resistance below 5 K at zero magnetic field was failed.
The MR is at least $1.2\times10^4$ in 9 T at 2 K, while it drops to less than 10 above 20 K.
The large MR for cadmium and zinc has been reported in Ref. \cite{Cd_Zn_in_Cdfor_MR_hall}, but we have a polycrystalline sample in this measurement.
Our measurement indicates that the large parabolic MR might be commonly observed for the compensated semimetals with large $\mathrm{RRR}$, and the resonance effect is a classical, two-band effect regardless the band structure in detail.

Finally we discuss the saturation of the MR for the compensated semimetals.
The parabolic MR for WTe$_2$ does not show any sign of saturation in a magnetic field up to 60 T, which is believed to be due to a prefect balance of the electron and hole pockets \cite{WTe2Cava, WTe2PRLforMobility}.
As far as we know, the saturations of the transversal MR for semimetals have been observed for bismuth\cite{BismuthHighfield} and graphite\cite{Graphite_high_field} which have small quantum limit less than 10 T.
On the other hand, the MR for our TaAs$_2$ intends to be saturated in 45 T while its main frequency of SdH oscillations is 50 T.
The derivation from a perfect $n-p$ balance is a plausible explanation for the saturation of the parabolic MR, but we notice that the saturation fields are close to the quantum limits for three semimetals.
The measurements for other semimetals in an intense magnetic field will help to understand this feature.

\section{conclusion}

To conclude, we grew large single crystals of NbAs$_2$ and TaAs$_2$ via a chemical vapor transport (CVT) process.
These two compounds are compensated semimetals with large, parabolic-field-dependent MR at low temperatures.
We believe that most of the large, parabolic MR for the compensated semimetals can be explained as a result of electron-hole balance.

\section{Acknowledgement}
Note : We notice several related works in Arxiv \cite{TA2_1, TA2_2, TA2_3} when we prepared this draft.

We thank Cheng-Long Zhang for pointing out the current jet problem in the negative longitudinal magnetoresistance.
We thanks Yuan Li and Ji Feng for using their instruments.
This project is supported by National Basic Research Program of China (GrantNos. 2013CB921901 and 2014CB239302).

\section{reference}

\bibliography{NA201_2}
\bibliographystyle{unsrt}

\begin{table}[ht]
\begin{flushleft}
\caption{\label{table1} RRR and MR for different samples of TaAs$_2$ and NbAs$_2$ grown by using different transport agents}
\begin{ruledtabular}  
\begin{tabular}[t]{llll}  
Sample & transport agent & RRR & MR at 2 K in 9 T\\  
\colrule  
N2 & I$_2$ & $53.6$ & $332$ \\
 \colrule
N5 & NbI$_5$: 75 mg & $317$ & $8.8\times10^3$ \\
\cline{1-1} \cline{3-4}
N9 &  & $313$ & $7.4\times10^3$ \\
\colrule
T1 & I$_2$  & $24.0$ & $95.7$\\
 \colrule
T2 & TaBr$_5$: 75 mg& $224$ & $3.2\times10^3$\\
\cline{1-1} \cline{3-4}
T3 & & $198$ & $2.0\times10^3$\\
\cline{1-1} \cline{3-4}
T4 & & $270$ & $5.8\times10^3$\\
\cline{1-1} \cline{3-4}
T17 & & $218$ & $3.6\times10^3$\\
 \colrule
T5 & TaBr$_5$: 74 mg & 44.0 & 315\\
 \cline{1-1} \cline{3-4}
T6 & and TeBr$_4$ : 4.7mg & 42.4 & 285\\
 \cline{1-1} \cline{3-4}
T7 & & 41.5 & 268\\
 \colrule
T8 & TaBr$_5$: 74 mg & 15.6 & 44.4\\
\cline{1-1} \cline{3-4}
T9 & and TeBr$_4$: 11.4mg & 17.3 & 53.8\\
 \cline{1-1} \cline{3-4}
T10 & & 18.6 & 55.1\\
 \colrule
T11 & TaBr$_5$: 74 mg & 33.0 & 168\\
\cline{1-1} \cline{3-4}
T13 & and TeBr$_4$: 22.3 mg & 38.4 & 208\\
 \colrule
T14 & TeBr$_4$: 35 mg & 2.30 & 0.093\\
\cline{1-1} \cline{3-4}
T15 & & 2.33 & 0.064\\
\cline{1-1} \cline{3-4}
T16 & & 2.42 & 0.258 \\
 \colrule

\colrule
\end{tabular}
\end{ruledtabular}All the samples come from 9 batches with different transfer agents. The growth conditions are the same as described in Experiment part.
\end{flushleft}
\end{table}

\begin{table}[ht]
\begin{flushleft}
\caption{\label{table2} Summarized parameters for the semimetals showing large parabolic MR at low temperatures. }
\begin{ruledtabular}  
\begin{tabular}[t]{llllll}  
semimetal  & $\mu$ (cm$^2$V$^{-1}$s$^{-1}$) & RRR & MR & reference\\
\colrule
Bi & N.A & 476 & $1.6\times10^7$ at 4.2 K 5 T & \cite{Bismuthlowtempnum1}\\
\colrule
Bi & $1.1\times10^7$ & 150 & $2\times10^4$ at 2 K 0.5 T & \cite{BismuthvalleyNatPhy}\\
\colrule
Bi & $3\times10^6$ & 40 & $1.1\times10^4$ at 4.2 K and 1.8 T & \cite{Bismuth_first_paper_in_PRL}\\
\colrule
Graphite & $1.0\times10^6$ & 37 & $8.3\times10^3$ at 4.2 K and 2.3 T & \cite{Graphite_useful_2}\\
\colrule
WTe$_2$ & $1\times10^5$ & 1256 & $3.1\times10^3$ at 1.2 K and 10 T & \cite{WTe2PRLforMobility}\\
\colrule
WTe$_2$ & $5\times10^4$ & 370 & $4.5\times10^3$ at 4.5 K and 14.7 T &  \cite{WTe2Cava}\\
\colrule
TaAs$_2$ & $2.1\times10^5$ & 270 & $5.8\times10^3$ at 2 K and 9 T &  current paper\\
\colrule
NbAs$_2$ & $1.6\times10^5$ & 317 & $8.8\times10^3$ at 2 K and 9 T &  current paper\\
\colrule
NbSb$_2$ & N.A & $450$ & $8.8\times10^3$ at 2 K and 9 T & \cite{NbSb2SciRep}\\
\colrule
Cd & N.A & $4.0\times10^4$ &  $1\times10^5$ at 1.4 K and 2.5 T & \cite{Cd_Zn_in_Cdfor_MR_hall}\\
\colrule
Cd (polycrystal) & N.A & $3\times10^5$ &  $1.2\times10^5$ at 2 K and 9 T & current paper \\
\end{tabular}
\end{ruledtabular}
The mobility data were fitted by two-carrier in the literatures.
\end{flushleft}
\end{table}

\end{document}